\pdfoutput=1 
\documentclass{JINST}
\usepackage{booktabs}
\usepackage{subfigure} 
\usepackage{url} 
\usepackage{lineno} 
\preprint{ATL-COM-INDET-2014-090}
\title{Experience on 3D Silicon Sensors for ATLAS IBL}

\author{G. Darbo$^a$\thanks{Corresponding author.} 
~ on behalf of the ATLAS Collaboration\\
\llap{$^a$}Istituto Nazionale di Fisica Nucleare - Sezione di Genova,\\
 via Dodecaneso 33, 16145 Genova, Italy\\
E-mail: \email{giovanni.darbo@ge.infn.it}}

\abstract{3D silicon sensors, where plasma micro-machining is used to etch deep narrow apertures in the silicon substrate to form electrodes of PIN junctions, represent possible solutions for inner pixel layers of the tracking detectors in high energy physics experiments. This type of sensors has been developed for the  Insertable B-Layer (IBL), an additional pixel layer that has been installed in ATLAS during the present shutdown of the LHC collider at CERN. It is presented here the experience in designing, testing and qualifying sensors and detector modules that have been used to equip part of the IBL. Based on the gained experience with 3D silicon sensors for the ATLAS IBL, we discuss possible new developments for the upgrade of ATLAS and CMS at the high-luminosity LHC (HL-LHC).}

\begin{document}
%
\section{Introduction}\label{Intro:1}
\begin{figure}[tbp] 
	\centering
	\includegraphics[width=0.85\textwidth]{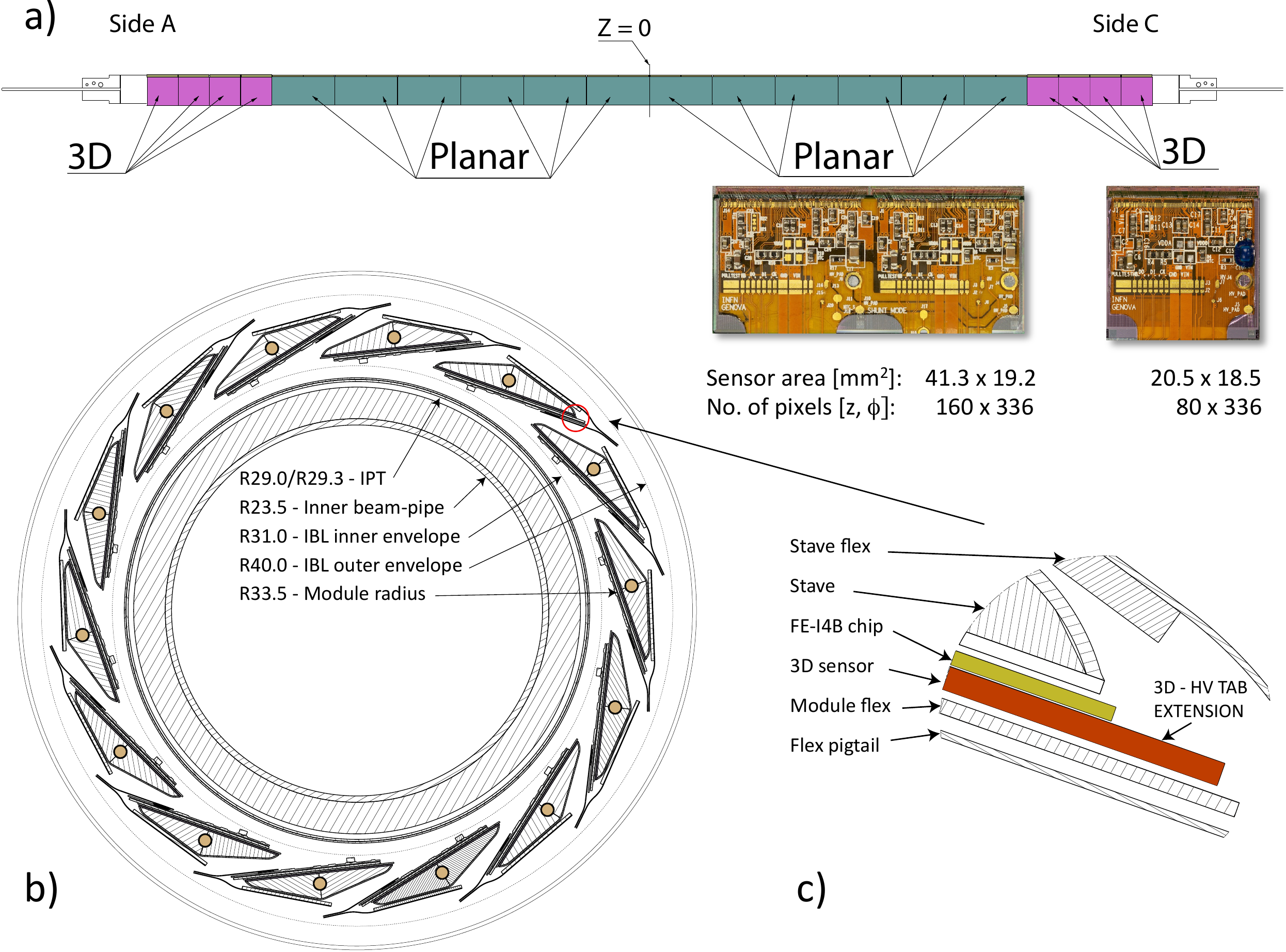}
	\caption{(a) Stave layout with the organization of planar and 3D sensor modules. (b) Layout of the IBL detector with the 14 staves around the IBL positioning tube (IPT) and (c) zoom of one stave side where a 3D sensor module is visibile.}
	\label{fig:fig_1}
\end{figure}

The Insertable B-Layer (IBL) \cite{IBL-TDR} is a fourth pixel layer added to the present Pixel Detector of the ATLAS experiment \cite{Aad:2008} at the Large Hadron Collider (LHC), between a new vacuum pipe and the current inner pixel layer. The principal motivations of the IBL are to provide track pattern recognition robustness, precision for b-tagging and vertexing performance as the instantaneous luminosity of the LHC increases beyond the design luminosity of $10^{34}  \rm{cm^{-2}s^{-1}}$ and the integrated radiation deteriorates the performance of innermost pixel layers. The requirements on the radiation damage  for the IBL was set to $5\times 10^{15} ~\rm{1\,MeV\,n_{eq}cm^{-2}}$ of NIEL (non-ionizing energy loss) and $2.5\,\rm{MGy}$ of TID (total ionizing dose).\\
3D silicon sensors, originally proposed in 1997 \cite{Parker:1997aa}, are suitable for high doses of nuclear interacting particles because their electrode distance is typically shorter than in usual planar sensors. In the 3D sensors the electrical field is defined by the column distance, while in the planar ones by the distance between the facing surfaces of the detector. After heavy irradiation damage, there is an increase of the full depletion voltage, and a decrease of the charge collection efficiency due to carrier trapping. Reducing the distance between electrodes lowers the depletion voltage: for 3D at IBL lifetime NIEL dose, less than 200\,V are needed to fully deplete the sensor, while a $200\,\rm{\mu m}$ thick planar requires 1000\,V or more.\\
In 2007 ATLAS launched an internal effort to study the potential of 3D sensors for future upgrades towards the high-luminosity LHC (HL-LHC); the ATLAS 3D Collaboration was then formed \cite{C.-Da-Via:2007aa}. In 2009 the IBL project has been started by ATLAS, with a plan of installation in 2016; the 3D collaboration decided to  prototype sensors fulfilling the requirement for the IBL in terms of radiation resistance and with a layout matching the FE-I4 chip in design \cite{collaboration:2012ab}. In 2011 there was a schedule advancement of two years of the ("fast-track") IBL installation and the sensor technology was reviewed on the basis of the built module prototypes with planar and 3D sensors. The measured performance of the 3D module prototypes, before and after irradiation, was convincing to use them for the first time in an experiment. The proposal was of a mixed-sensor IBL layout: planar sensors in the central region and 3D in the forward/backward part, where tracking would benefit of a more uniform charge collection across the sensor depth after irradiation.
The IBL layout is shown in figure~\ref{fig:fig_1}. There are 14 staves in a turbine structure; each stave has 12 modules with double-chip planar sensors in the center and 4 forward single-chip 3D sensors at the two extremities.\\
As of today the IBL detector is completed, installed in ATLAS under commissioning and ready for the next year restarting of LHC.
\section{Sensor design, production and results}
\begin{figure}[ht!] 
\centering
\subfigure[a][]{\includegraphics[width=0.379\linewidth] {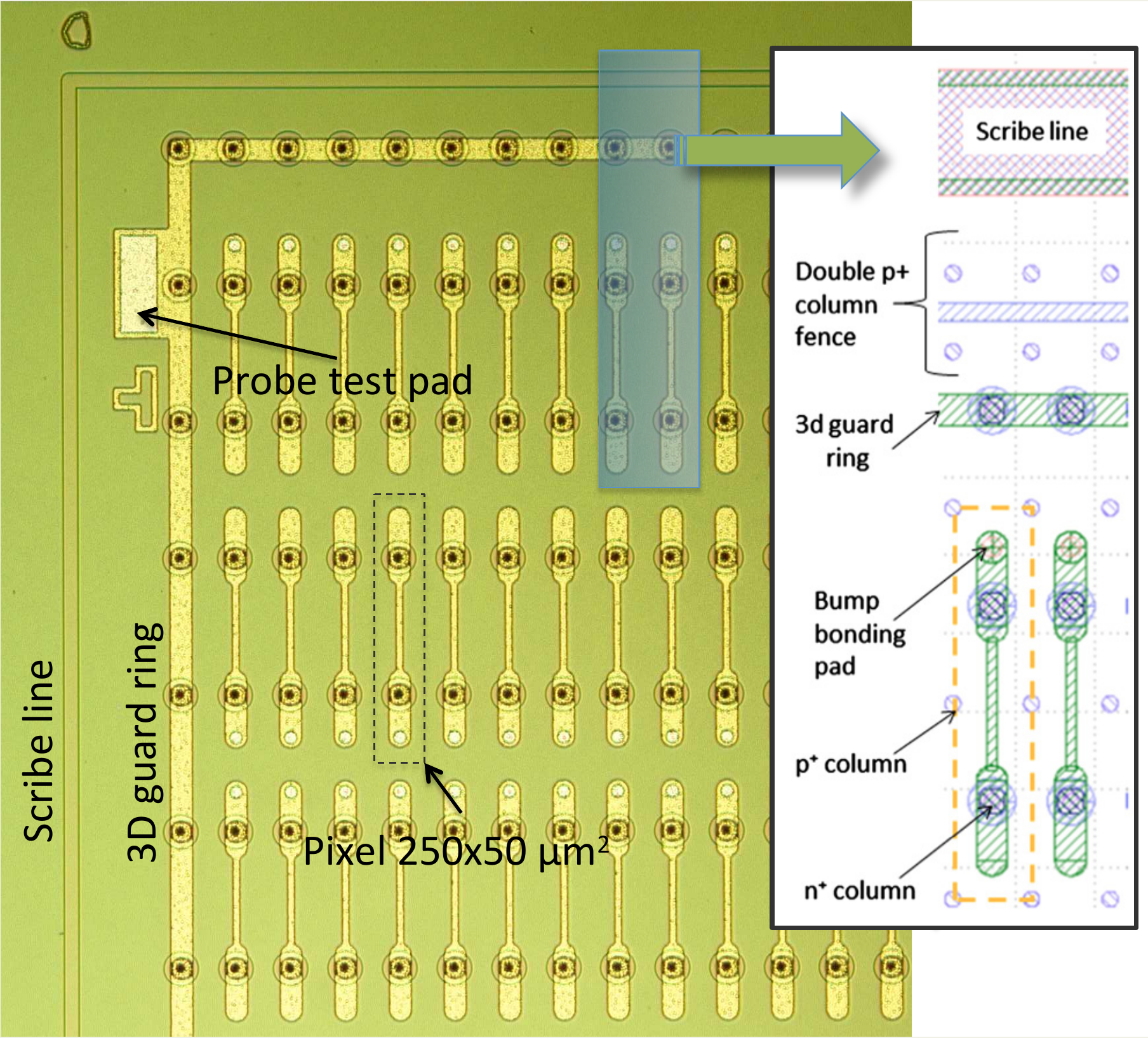} \label{fig:fig_3a}}
\hspace{0.5cm}
\subfigure[b][]{\includegraphics[width=0.380\linewidth] {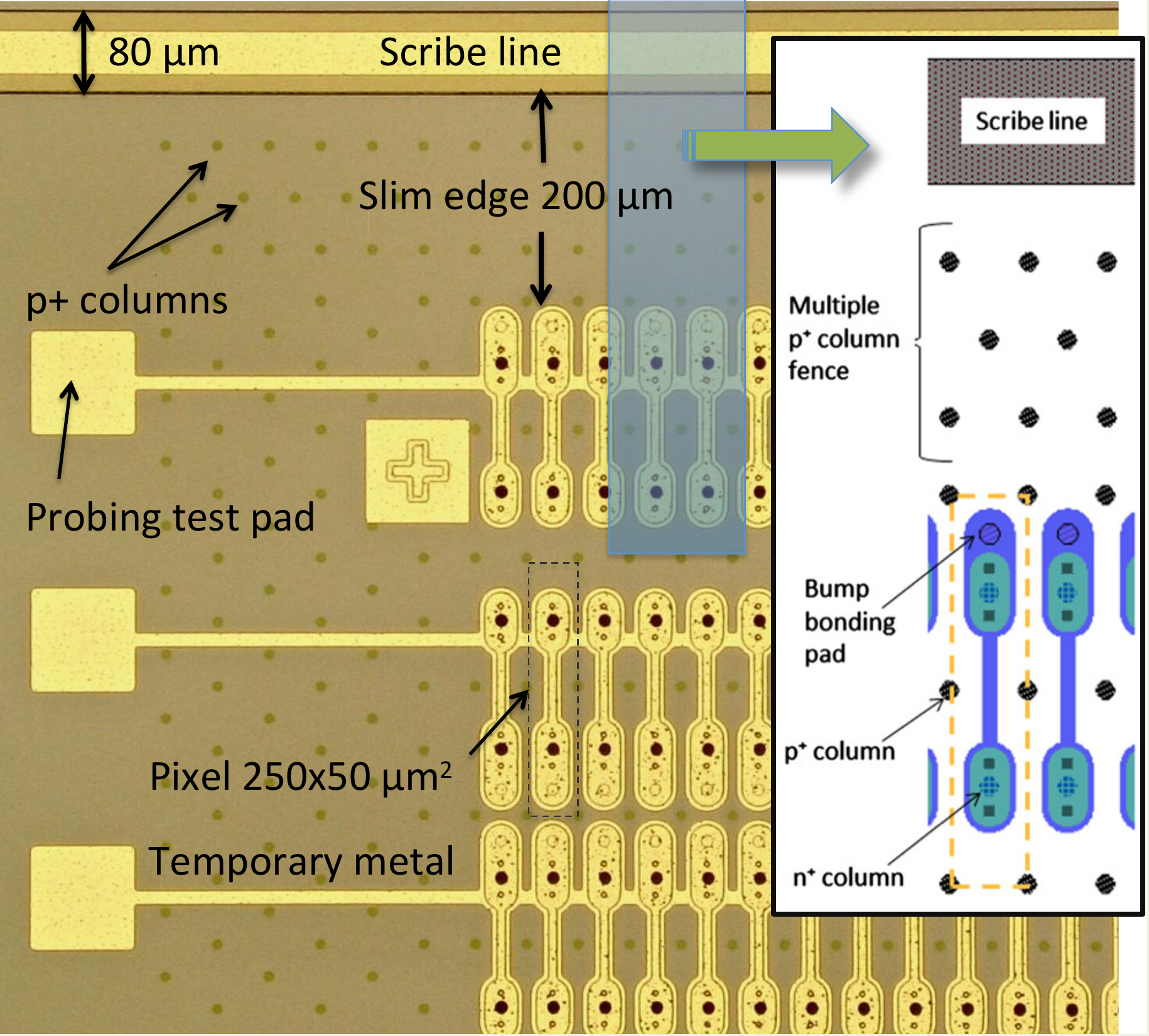} \label{fig:fig_3b}}
  \caption{Microphotograph of 3D sensors seen from front-side with blowup layout comparisons.The two photos show: (a) CNM and (b) FBK sensors for FE-I4 pixel chips.} 
  \label{fig:fig_3}
\end{figure}

\begin{figure}[th] 
\centering
\subfigure[a][]{\includegraphics[width=0.3824\linewidth] {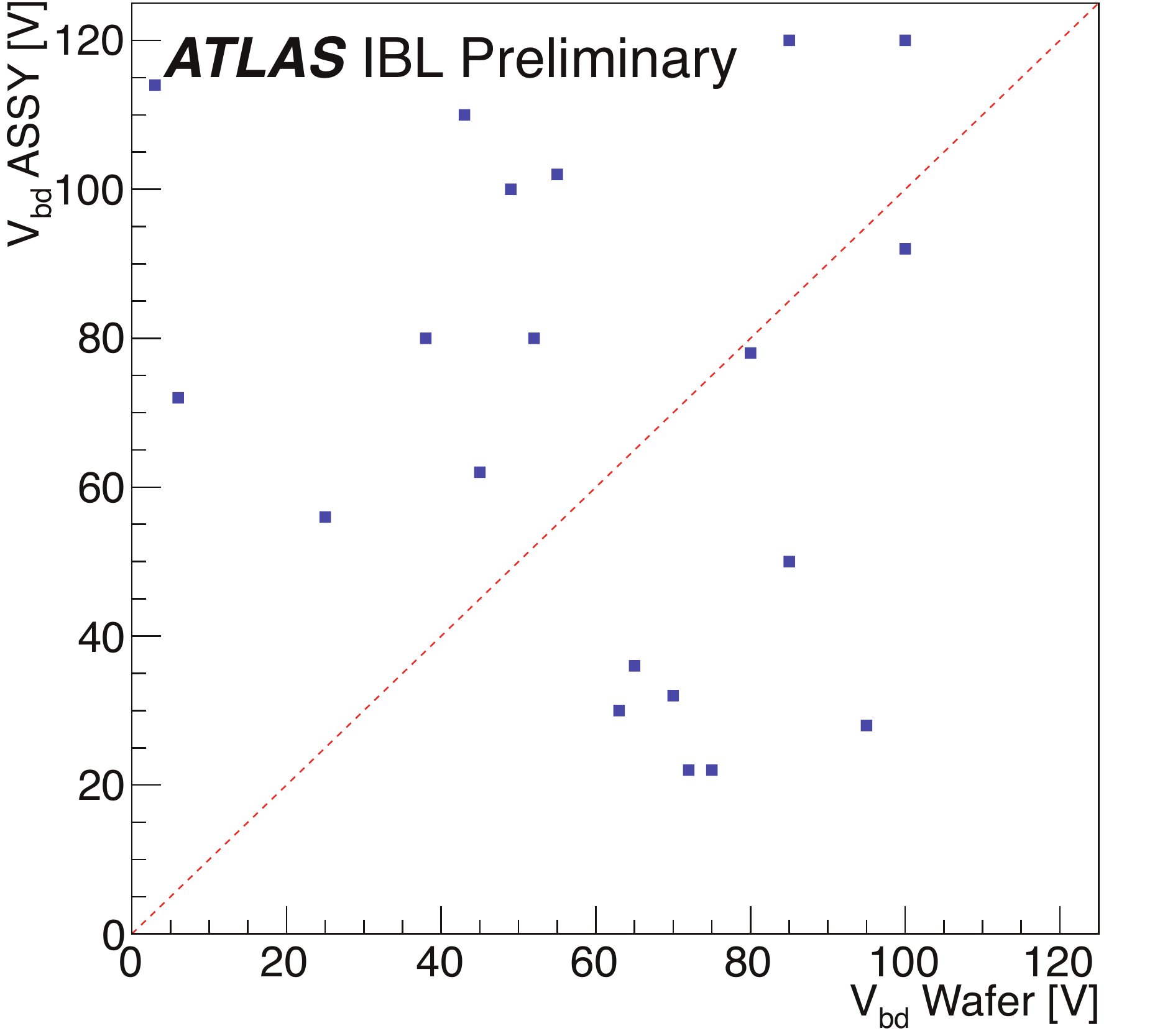} \label{fig:fig_4a}}
\subfigure[b][]{\includegraphics[width=0.4016\linewidth] {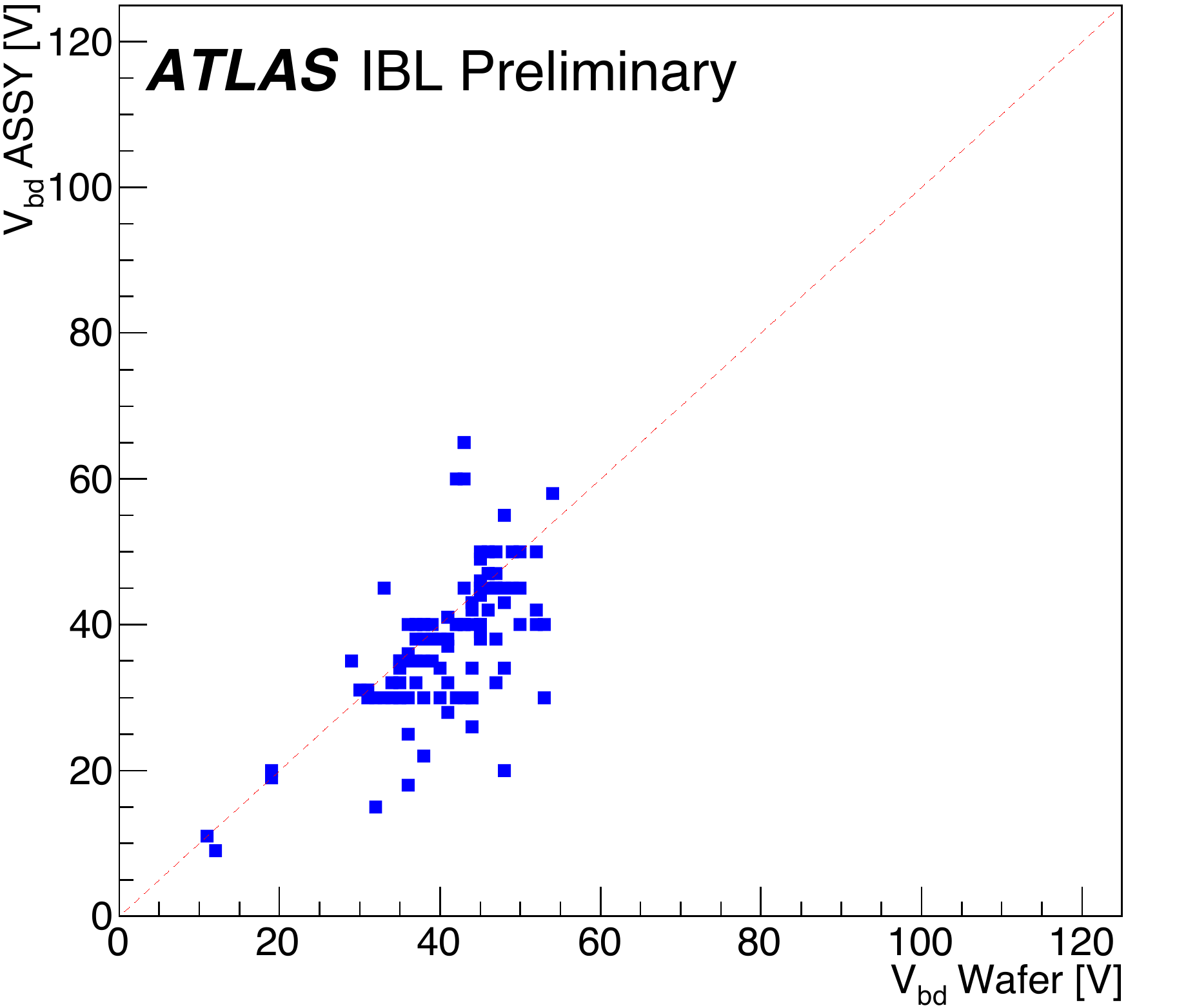} \label{fig:fig_4b}}
\caption{Measurement of sensor breakdown voltage ($\rm{V_{bd}}$) after this is assembled into a detector module (referred as "ASSY") versus  on-wafer. Measurement on wafer is done by contacting the fence guard ring for CMN (a) and by temporary metal connecting pixels into strips for FBK (b). Only a sub-sample of CNM tiles is shown in the plot, for modules where I--V curves were available with a current compliance of $\rm{100\,\mu A}$~\cite{PixelPlots}.}
\label{fig:fig_4}
\end{figure}

The 3D silicon sensors used in the IBL have been produced by two silicon foundries~\cite{Da-Via:2012aa,Pellegrini201327,Giacomini20132357}: CNM\footnote{Centro Nacional de Microelectronica, CNM-IMB (CSIC), Barcelona E-08193, Spain} and FBK\footnote{Fondazione Bruno Kessler, FBK-CMM, Via Sommarive 18, I-38123 Trento, Italy}, on $\rm{230\,\mu m}$ thick 4-inch FZ\footnote{Silicon crystal growth methods: FZ -- float zone; CZ -- Czochralski} p-type wafers having a resistivity of $10-30~\rm{k}\Omega\,\rm{cm}$. A wafer floorplan and sensor geometry for FE-I4~\cite{collaboration:2012ab} pixel front-end chip was defined in common with the different sensor producers participating in the prototype program coordinated by the ATLAS 3D Collaboration. A total of 8 FE-I4 single-chip sensors fits in a wafer layout.
In addition to the two already mentioned foundries also SINTEF\footnote{SINTEF MiNaLab, Blindern, N-0314 Oslo, Norway} and SNF\footnote{Stanford Nanofabrication Facility, Stanford, CA, United States} participated in the prototype program. 
\begin{figure}[tbp] 
	\centering
	\includegraphics[width=0.8\textwidth]{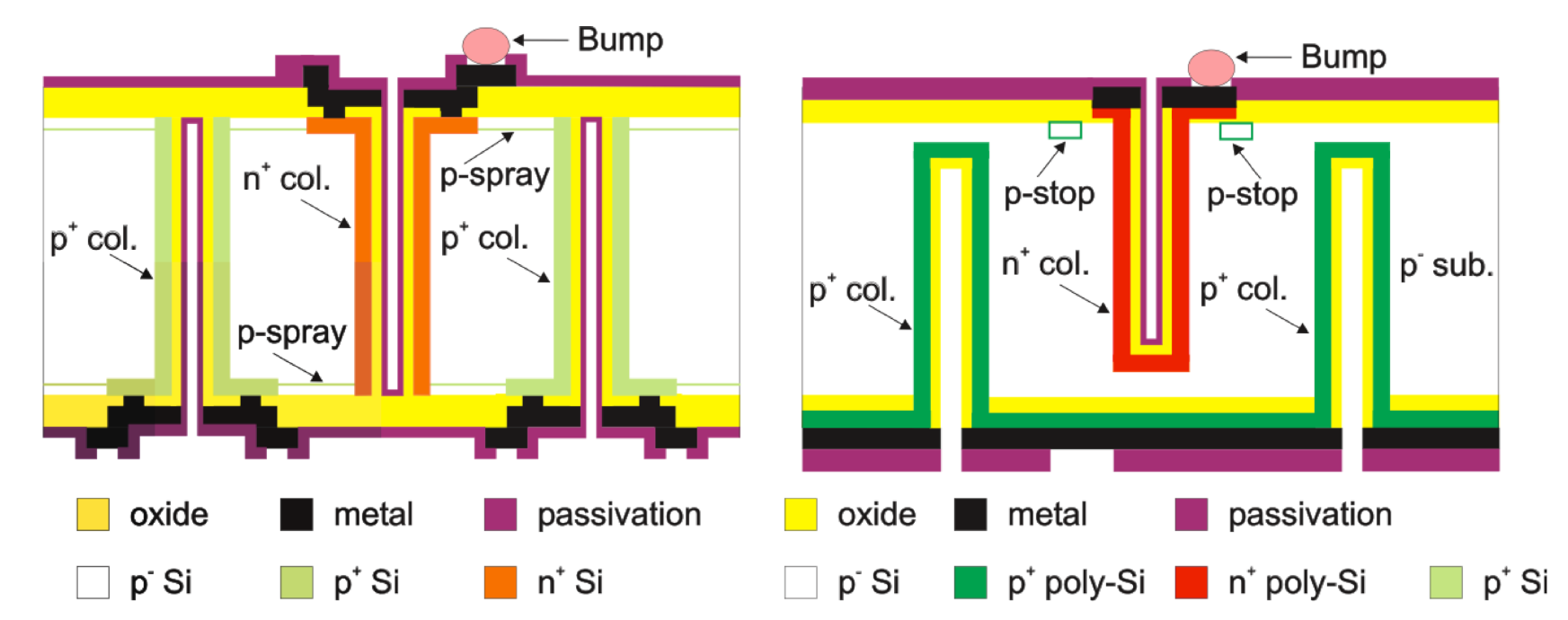}
	\caption{Schematic cross-section of the 3D detector with passing-through columns from FBK (left) and with partial columns from CNM (right) fabricated on a p-type substrate (not to scale)~\cite{collaboration:2012ab}. }
	\label{fig:fig_2}
\end{figure}
\paragraph{Layout and process.} The schematic cross-section for the 3D sensors used by FBK and CNM are shown in figure~\ref{fig:fig_2}, while figure~\ref{fig:fig_3}  shows micro-photographies of a corner of the sensors together with layout blowups. In both cases a double-sided process is used, with $\rm{n^+}$-columns (junction) etched from the front wafer side (bump-bonding side) and $\rm{p^+}$-columns (ohmic) from the back side. 
In the CNM processing, columns do not pass all through the wafer, but stop at a short distance from the surface of the opposite side~\cite{Pellegrini201327}; in the case of FBK sensors the original technology, similar to CNM, was later modified for the IBL to allow for passing through columns~\cite{Giacomini20132357}. Additional difference in the process of the columns is the partial filling with poly-silicon in case of CNM, while FBK leaves them empty.\\
SINTEF and SNF use a  single-sided process with both junction and ohmic columns etched from the front side. This process allows for active sensor edges by the use of etched trenches completely filled with $\rm{p^+}$ doped poly-silicon. Such technology requires a handler wafer oxide-bonded to the device wafer, which needs extra steps to attach and remove.
Single-sided 3D sensors have the bias connection on the front side, which is bump-bonded to the read-out chip. To apply the bias it is therefore necessary to extend the sensor tile with a tab overhanging from the front-end chip as shown in figure~\ref{fig:fig_1}~\cite{Da-Via:2012aa}. To keep a common floorplan also CNM and FBK sensors have such extension tab, even if it is not used. It could have been removed for IBL sensor production, but the fast track IBL schedule
did not allow for a redesign of the photolithography masks.\\  
CNM and FBK designs have a $200\,\rm{\mu m}$ slim edge. To prevent the currents generated by the dicing defects in the crystal from reaching the active area, an edge termination structure is placed all around the sensor tile. In case of FBK, the termination is made of all ohmic columns biased at the substrate voltage (see figure~\ref{fig:fig_3b}), whereas for CNM, in addition to ohmic columns, a 3D guard ring is used, which is grounded throughout the front-end chip by dedicated bump-bond pads (see figure~\ref{fig:fig_3a}). It has been successively shown that the FBK $200\,\rm{\mu m}$ edge can be reduced to less than $100\,\rm{\mu m}$ without significantly affecting sensor performance, i.e., leakage current and breakdown voltage~\cite{Povoli:2012aa}.\\
For FBK sensors, surface isolation in between junction columns is ensured by a p-spray layer (furthermore, being the columns full passing, the p-spray is necessary on both wafer sides); while for CNM sensor this is done by p-stop.
\paragraph{Test on wafer and production yield.} To test sensor tiles at wafer level, FBK (and also SINTEF/SNF) uses a temporary metal shown in figure~\ref{fig:fig_3b}. The deposited metal connects all the 336 pixels in one column transforming the pixel sensor into a strip detector with probing pads on one tile side. This aluminum layer is deposited at the end of the fabrication process and removed after the test. The I--V of each of the 80 strips is measured by a dedicated probe card with respect to the back-side metallization. Each I--V is representative of all the 336 pixels shorted together; the total I--V is obtained by summing up the 80 curves. \\
For CNM sensors, the temporary metal was not fully compatible with the process and available testing instruments at the foundry and a different testing strategy was used. In this case, the guard ring (see figure~\ref{fig:fig_3a}) current is evaluated as a function of the applied voltage.\\ 
Tiles that have breakdown voltage ($\rm{V_{bd}}$) greater than 25\,V and leakage current ($\rm{I_{leak}}$) at 20\,V respectively less than $\rm{0.2\,\mu A}$ for CNM guard ring and $\rm{2\,\mu A}$ for the whole tile for FBK are selected for assembly into full module, i.e., bumb-bonded to FE-I4 and dressed with the flex-hybrid circuit. The sensor I--V is re-measured on assembled modules. Correlation between $\rm{V_{bd}}$ measured on tile and on assembled module shows that for FBK the selection criterium is valid (see figure~\ref{fig:fig_4b}), whereas guard ring measurement is not good enough (see figure~\ref{fig:fig_4a}). CNM is considering alternative testing methods for future 3D sensor designs, like a high resistivity poly-silicon biasing grid.\\
Out of all processed and tested wafers, there are 33 wafers from FBK and 40 from CNM that passed the selection criteria of having three or more good tiles (basically fulfilling $\rm{V_{bd}}$, $\rm{I_{leak}}$ and mechanical quality requirements). Selected wafers are then processed for under bump metallization (UBM), needed step for bump-bonding at IZM\footnote{Fraunhofer IZM-Berlin, Gustav-Meyer-Allee 25, 13355 Berlin.}, and good tiles are flip-chipped onto FE-I4. All modules received from IZM are assembled adding a flex-hybrid circuit, glued on top of the sensor back-side, that is then wire-bonded to the FE-I4 for signal and power and to the sensor for bias voltage. The yield for the module assembly is $50\,\%$ (i.e. 84 modules) for CNM and $56\,\%$ (i.e. 66 modules) for FBK considering all batches from IZM. Major yield killer are disconnected bumps for FBK and the breakdown voltage for CNM (see the guard ring criterium).  The bump-bonding process has been cured after the first production batch and the yield for remaining part of the modules increased to $63\,\%$ for CNM and to $62\,\%$ for FBK.
\section{Experience with sensor: performance with modules and in the IBL}
\begin{figure}[t!] 
\centering
\subfigure[a][]{\includegraphics[width=0.46\linewidth] {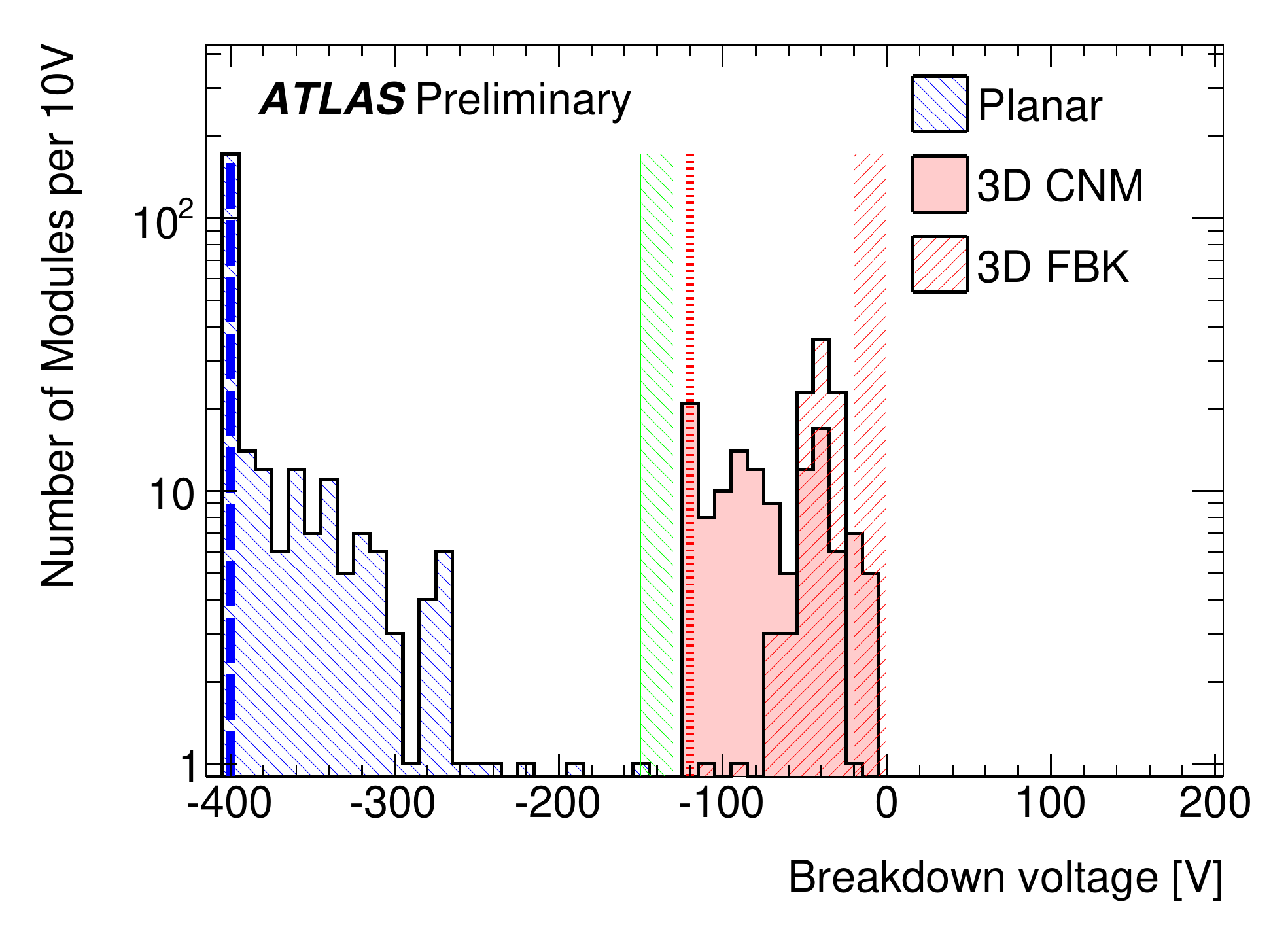} \label{fig:fig_5a}}
\subfigure[b][]{\includegraphics[width=0.48\linewidth] {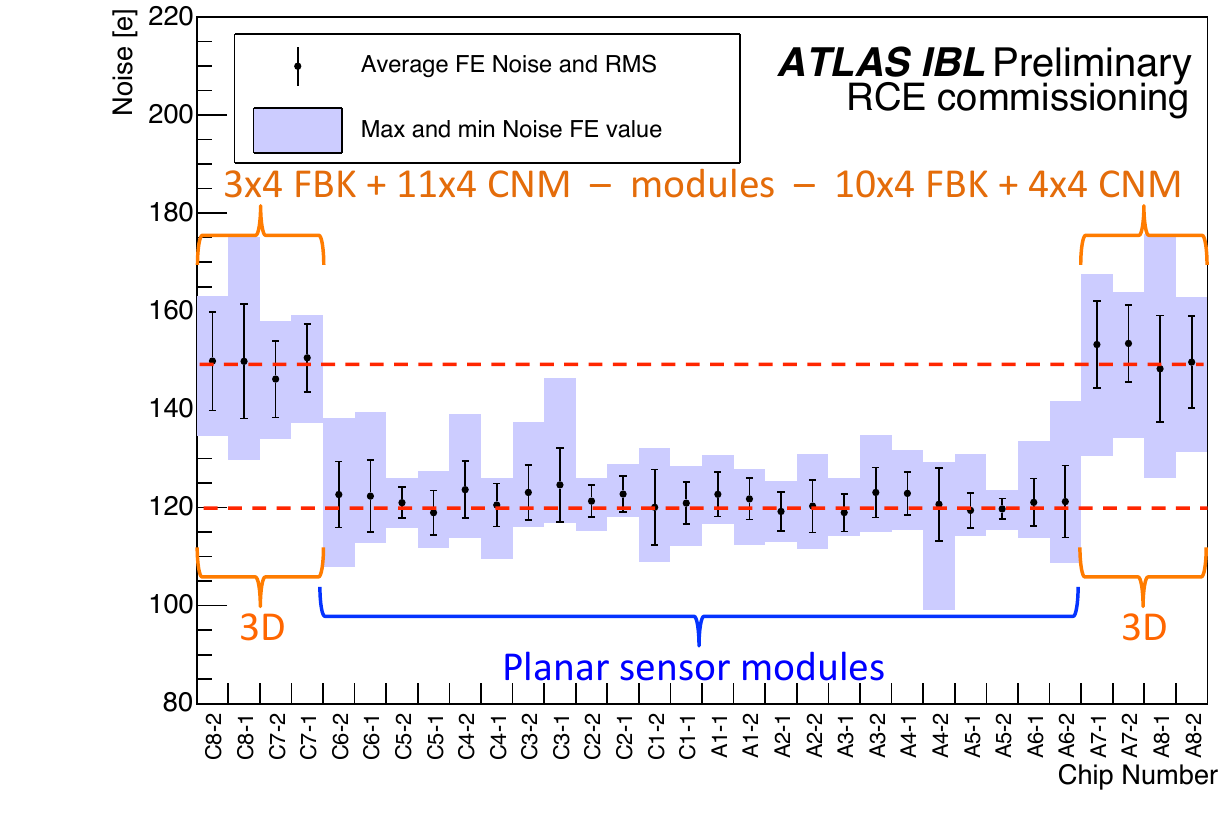} \label{fig:fig_5b}}
  \caption{Distribution of $\rm{V_{bd}}$ for the whole production of IBL modules (a) and average modules noise for all the 14 IBL staves during commissioning after installation (b). The vertical bars (a) represent the minimum accepted $\rm{V_{bd}}$ for 3D (red) and planar (green) sensors by the IBL quality assurance~\cite{PixelPlots}.}
  \label{fig:fig_5}
\end{figure}

Sensor and modules have been extensively studied at test-beam and in the laboratory, prior and after irradiation, as part of the quality assurance (QA) of  modules and loaded staves, and finally in the commissioning of the IBL detector after installation in ATLAS. Some results are given here.
\paragraph{Depletion and breakdown voltage.} The sensor depletion voltage is typically lower for FBK  than for CNM sensors. This is mainly due to the different full-through versus partial column designs. On the other hand the breakdown voltage ($\rm{V_{bd}}$) for CNM sensors is significantly higher before irradiation (see figure~\ref{fig:fig_5a}), and marginally higher after IBL lifetime dose ($\rm{5\times 10^{15}\,~n_{eq}\, cm^{-1}}$). 
\paragraph{Capacitance and noise.} Modules with 3D sensors have been extensively operated with threshold as low as 1500\,$\rm{e^-}$ at the test beam, irradiated and non, and as part of the QA of modules on local supports (staves). Noise is slightly  higher for FBK (mean =140.3\,$\rm{e^-}$) respect to CNM (mean =130.7\,$\rm{e^-}$) as measured (discriminator threshold set at 3000\,$\rm{e^-}$ and operating at a temperature of approximately $\rm{-15^{\circ}C})$ in the entire set of modules passing the IBL QA. Higher noise in the FBK modules is due to the higher capacitance of the pixel (all through columns). The noise measured on installed IBL, with final cabling, is slightly higher than what measured on the module QA. In figure~\ref{fig:fig_5b} the distribution of average module noise for all the modules in the IBL is shown. Planar modules have an average noise that is 30\,$\rm{e^-}$ lower that 3D. This is compatible with a lower measured capacitance of planar pixel sensors of 110\,fF respect to 169\,fF of 3D~\cite{Havranek:2013aa}.
\paragraph{Track efficiency.} Efficiency of 3D sensors has been evaluated at several test beam campaigns. Un-irradiated 3D sensors have shown nearly 100\,\% efficiency when tilted by $15^{\circ}$; at $0^{\circ}$ CNM sensors have shown $99.6\,\%$ efficiency and FBK $98.8\,\%$. Higher efficiency of CNM sensors is explained by having shorter columns and  being the charge under junction and ohmic columns contributing to the total signal. After an irradiation dose of $\rm{5\times 10^{15}\,\,n_{eq}\, cm^{-1}}$, the efficiency reaches 99.0\,\% for CNM and 98.2\,\% for FBK sensors, with modules tilted at $15^{\circ}$ and operating at $\rm{V_{bias}}$ of $\rm{160\, V}$.  Efficiencies quoted here have been measured with modules operating at a threshold of $\rm{1500-1600}$\,$\rm{e^-}$.
%
\begin{figure}[tbp] 
\centering
\includegraphics[width=0.73\textwidth]{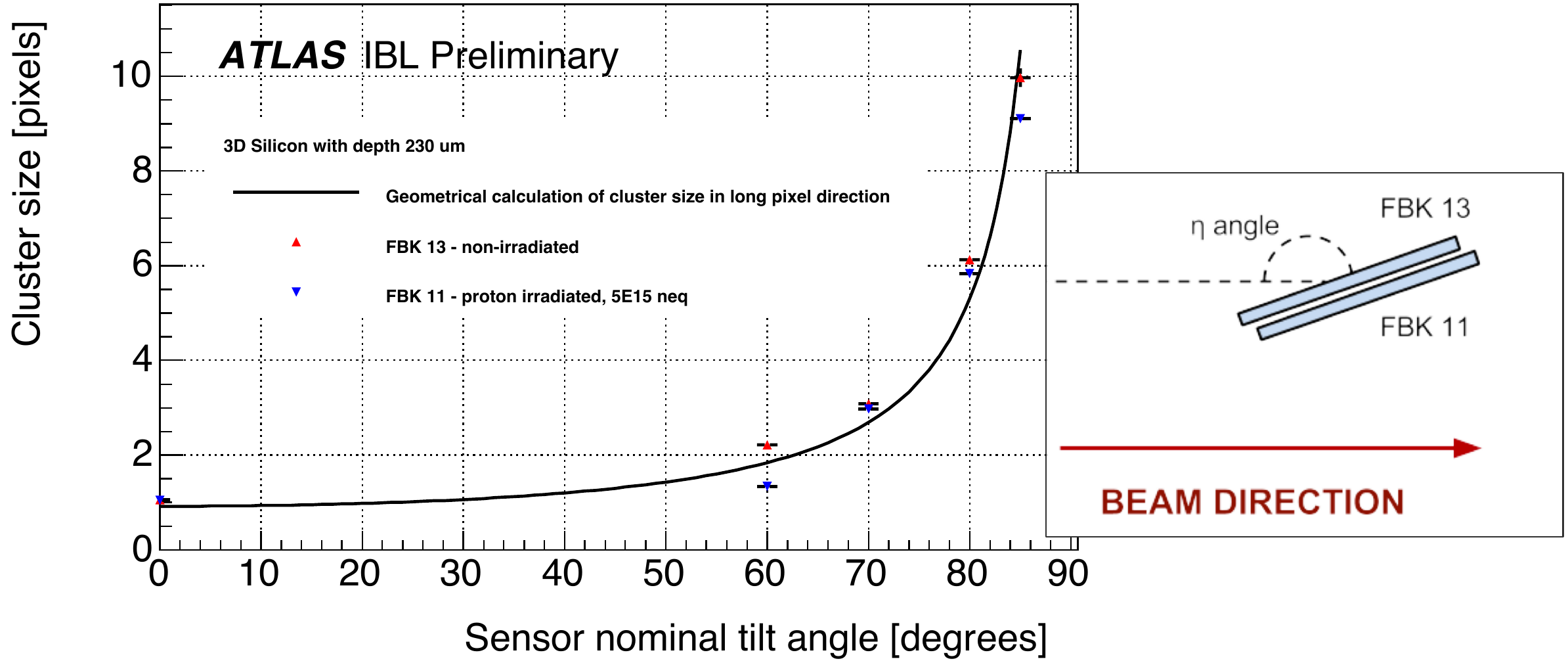}
\caption{Cluster	size as function of incidence angle in the long pixel direction. The comparison is between two 3D FBK modules, one irradiated to $\rm{5\times 10^{15} n_{eq}\,cm^{-2}}$ and the other un-irradiated~\cite{Nellist:2013aa}.}
\label{fig:fig_6}
\end{figure}
\paragraph{Small incidence angle behavior.} 3D modules are located at both extremities of the IBL, where interaction tracks are at small incidence angle. For precision tracking the uniformity of the charge collection across the sensor thickness is very important: small variations in cluster size affect the position of the hit. Test-beam measurements have been dedicated to characterize modules that have been tilted in the direction of the long pixels, simulating tracks at grazing angle as expected in the IBL forward regions. Figure~\ref{fig:fig_6} shows the measured and geometrically calculated cluster sizes as a function of the tilt angle. Un-irradiated and irradiated 3D sensors show the same size of cluster, which also reproduce the value predicted geometrically; this is an indication that charge is collected uniformly across the pixel depth~\cite{Nellist:2013aa}.
\begin{figure}[hbt!] 
\centering
\subfigure[a][]{\includegraphics[width=0.45\linewidth] {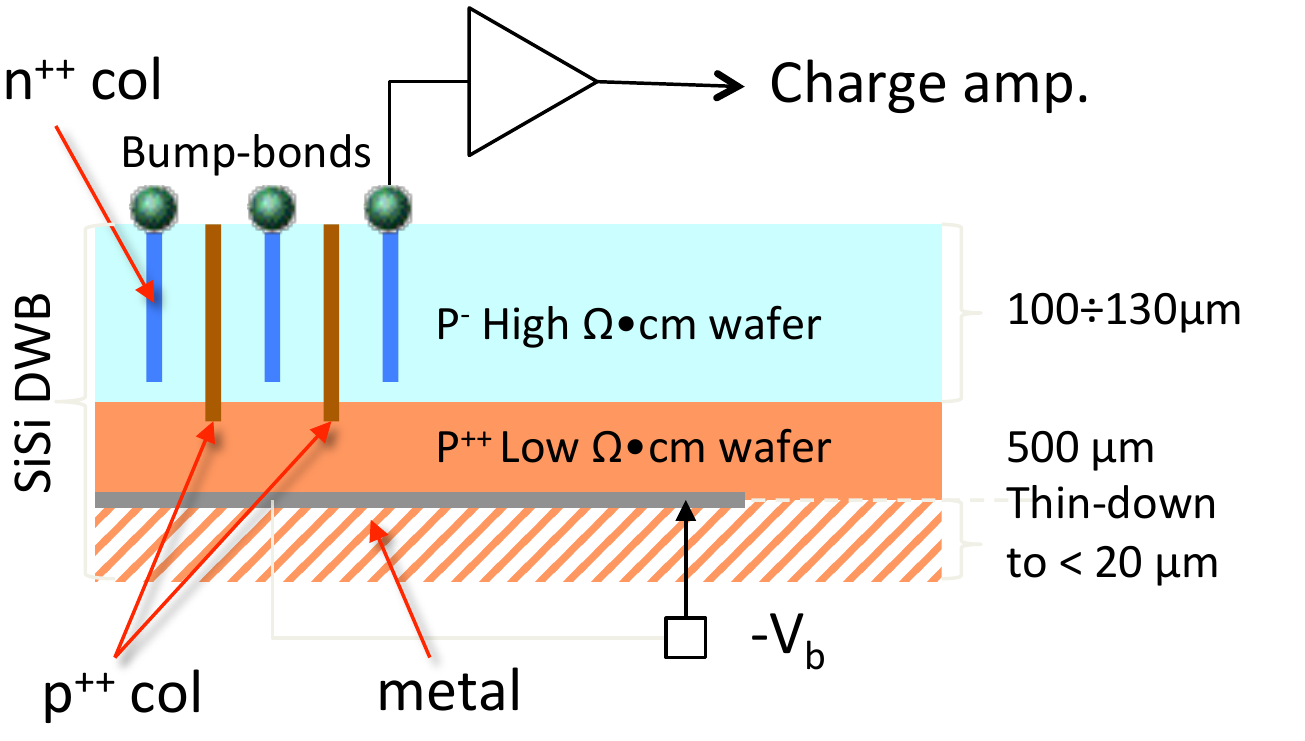} \label{fig:fig_7a}}
\subfigure[b][]{\includegraphics[width=0.27\linewidth] {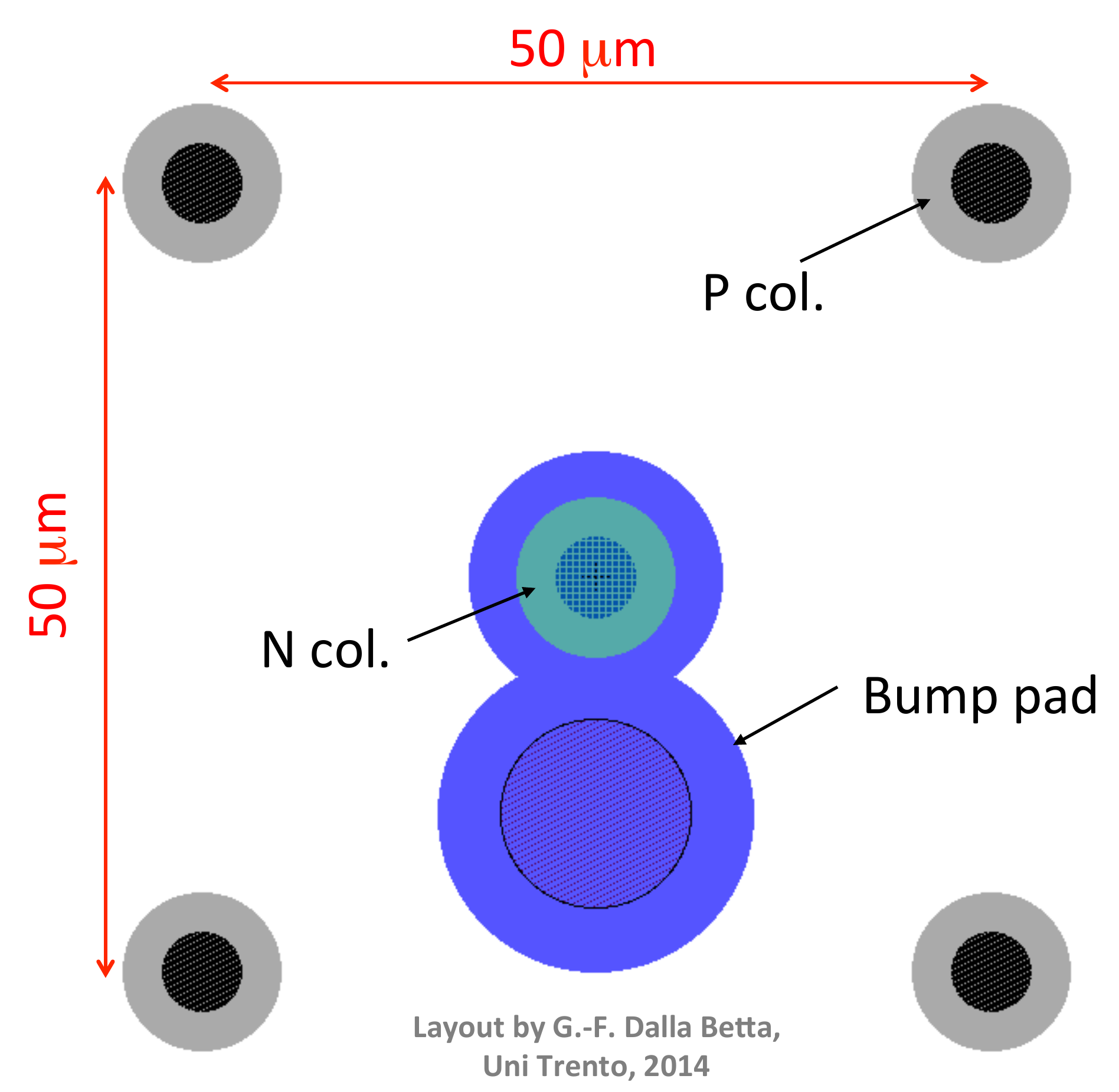} \label{fig:fig_7b}}
  \caption{Illustration of the new 3D process at FBK on SiSi direct wafer bond substrates (a) with layout of the pixel cell (b), courtesy of G.F. Dalla Betta~\cite{GFprivate}.}
  \label{fig:fig_7}
\end{figure}
\section{Looking at the future: 3D sensors for experiments upgrade at HL-LHC}
A new generation of 3D sensors is under development for the upgrades of ATLAS and CMS at the HL-LHC.  The new pixel detectors will reach four times more dose ($\rm{2\times 10^{16}\,n_{eq}\, cm^{-1}}$) in their innermost layers and will have five times smaller pixels ($\rm{50\times 50\,\mu m^2}$ or $\rm{25\times 100\,\mu m^2}$).  To cope with such requirements, the 3D sensors need some improvements. Smaller pixel area calls for thinner sensors; higher radiation dose asks for the reduction of the electrode spacing to overcome charge trapping. Pixel capacitance has to go down to reduce power and noise in the pixel electronics. The signal expected for 3D pixel cells, from semi-analytical model~\cite{DaVia:2009aa}, in $\rm{150\,\mu m}$ thick detector after total HL-LHC dose is in the range of $5000-5700\,\rm{e^-}$, being the lower limit for square pixels ($\rm{50\times 50\,\mu m^2}$) with one read-out electrode (see layout in figure~\ref{fig:fig_7b}) and the higher limit for rectangular ($\rm{25\times 100\,\mu m^2}$) with two read-out electrodes. Due to the number of electrodes, capacitance, instead, is higher for rectangular pixels than for square ones: $\rm{50-100\,fF}$ for a thickness of $\rm{150\,\mu m}$~\cite{GFprivate}.
Since both collected charge and signal efficiency decrease with reduced thickness and increased radiation dose, lower threshold operation becomes necessary. Again noise is an important factor to look at for low threshold operation. To increase efficiency it is also important to reduce the column diameter: either go to shorter columns or increase the aspect ratio.
The 3D process is typically more complex than that of a planar process. In the IBL, the number of process steps (masks) was significantly bigger than that of the planar devices and also the yield was lower. Improvement of both is important: the new process that will be used by FBK has 30\,\% less steps and also criticality of the layout, causing early voltage breakdown, has been studied and improved. Another way to reduce cost is going to 6-inch wafers.\\
Figure~\ref{fig:fig_7a}  shows the new process in development at FBK. This process is single-sided and uses wafer bonded substrates: a low resistivity $\rm{CZ^3}$ wafer is directly bonded (without oxide interface) to a FZ device wafer having high resistivity. The low resistivity wafer is used as mechanical support and electrically conductive backplane to bring the bias voltage to the ohmic columns; the high resistivity wafer thickness is optimized for charge collection performance and not for mechanical reasons. Handling wafer is afterward thinned down by grinding process to a value good to guarantee enough mechanical robustness.
CNM is also looking for improvements in the process, as the previously mentioned poly-silicon grid for I--V testing on wafer, or as the increasing of the columns aspect ration using a cryogenic DRIE (deep reactive ionizing etching) process.   
\section{Conclusions} 
\vskip -0.2cm
3D sensors have been successfully developed and produced for the IBL detector; they are used for the first time in an experiment fulfilling  the requirements of the new ATLAS pixel layer. The experience gained from IBL has shown to be very useful to improve this type of detectors for the next generation of pixel trackers in ATLAS and CMS at the HL-LHC.
\acknowledgments
\vskip -0.25cm
The author wishes to thanks: M.~Backhaus, G.F.~Dalla Betta, A.~Gaudiello, I.~L\'opez Paz, B.~Mandelli, R. Mendicino, C.~Nellist, A.~Sciuccati for fruitful discussions and material provided for the talk and the proceedings. \\
This work has been supported by the agencies funding the IBL~\cite{collaboration:2012ab} and, partially, by the European Commission under the FP7 Research Infrastructures project AIDA, grant agreement no. 262025. 

\end{document}